\documentclass[12pt,oneside,final,a4paper]{article}

\title{ Quantum-like formalism for cognitive measurements }
\author{Andrei Khrennikov\\
International Center for Mathematical Modeling\\
in Physics and Cognitive Sciences,\\
MSI, University of V\"axj\"o, S-35195, Sweden\\
Email: Andrei.Khrennikov@msi.vxu.se}
\begin{document}
\maketitle

\begin{abstract} We develop a quantum formalism
(Hilbert space probabilistic calculus) for measurements 
performed over cognitive 
systems. In particular, this formalism is used for mathematical modeling
of the functioning of consciousness as a self-measuring quantum-like system.
By using this formalism we could predict averages
of cognitive observables. Reflecting the basic idea of
neurophysiological and psychological studies on a hierarchic
structure of cognitive processes, we use $p$-adic hierarchic 
trees as a mathematical model of a mental space. We also
briefly discuss the general problem of the choice of an adequate
mental geometry.
\end{abstract}

\section{Introduction}
Since the creation of quantum mechanics, there are continuous
discussions on possible connections between quantum and 
mental phenomena. During the last hundred years, 
a huge number of various proposals and speculations have been presented. We shall
mention just a few of them. 

The philosophic system of Whitehead [1]-[3]
was the first attempt to establish a quantum--mental (or more precisely mental $\to$ quantum)
connection. In Whitehead's philosophy of the organism ``quantum" was some feature of basic
protomental elements of reality, namely {\it actual occasions,} see [1], especially p. 401-403.
See also A. Shimony [4] for modern reconsideration of quantum counterpart of Whitehead's philosophy 
of organism. It is especially important for us to underline that all protomental elements of reality
have quantum temporal structure: {\small ``The actual entity is the enjoyment of a certain quantum 
of physical time."} -- [1], p. 401.

The extended discussion on quantum--mental connection
was induced by attempts to solve the problem of quantum measurements, see e.g. [5]-[12].
The most extreme point of view is that physical reality is, in fact, created by
acts of observations. This kind of considerations is especially
closely related to the so called orthodox Copenhagen interpretation of quantum mechanics.\footnote{We mention
Berkeley's idealism as one of sources for such a point of view to physical reality.}
By this interpretation a wave function provides the complete description
of an {\it individual} quantum system. An act of measurement induces 
collapse of the wave function. The problem of measurement is still unsolved within a quantum framework
(at least on the basis of the conventional interpretation of quantum mechanics, see also section 10).
Among various attempts to provide a reasonable explanation of wave function reduction, one should mention
attempts to use consciousness as the determining factor of reductions of wave functions, see e.g. Wigner [8]. 

There were also  various attempts reduce an act of thinking to  quantum collapse, 
see e.g. Orlov [13] (quantum logic of consciousness); see also
Penrose [14], [15]: {\small ``I am speculating that the action of conscious thinking 
is very much tied up with the resolving out of alternatives that were previously in linear superposition."}

In fact, Penrose  worked in the reductionist approach, see e.g. [16] (and compare e.g. [17]-[20]): 
It seems we could not reduce cognitive phenomena to the physical activity of neurons. It might be that we could reduce it 
to activity of quantum systems. Roughly  speaking an act of  thinking is reduced to the collapse
of a wave function in quantum gravity. Our thinking ability is based on collapses of 
superpositions of two mass states. 

The idea of quantum-physical reduction
for cognitive processes is quite popular within the quantum community. We also mention
the investigations of H. Stapp [21] who used the 
Copenhagen (Heisenberg-potentiality)  approach to quantum mechanics.
He also used a quantum reductionist approach:  {\small ``Brain processes involve chemical
processes and hence must, in principle, be treated
quantum mechanically."} We should also mention 
quantum field reductionist models, by Jibu and Yasue [22], [23] (based on Umezawa [24]), 
Vitiello et el. [25]. These quantum field models look
more attractive (al least for me). At the moment there is no idea  how  make the great jump from
individual gravitational collapses to global acts of cognition. Quantum field models
are more useful to provide such a global structure connecting individual quantum collapses to 
global acts of thinking.

However, it seems that reductionism as the general methodology of 
studying the brain  is less and less popular in cognitive sciences.
After the period of large hopes associated with new possibilities
to study neuronal firings, there is strong disillusionment in the possibility
of some physical reduction of mental processes. This is one reason for the quite
strong critical attitude against quantum models in cognitive sciences. In the 
extreme form this criticism is expressed in the following form: 

{\it "The only common thing between
quantum and mental is that we have no idea how to understand any of these phenomena."}

Another thing that induces prejudice against quantum-reduction theories among
neurophysiologists is that a quantum micro description contains many parameters
that are far from magnitudes of corresponding brain's parameters
(e.g. temperature, time scale and  so on). Thus creators of all 
quantum reductionist models of  brain's functioning become immediately involved in 
hard battles with these parameters (e.g. high temperature of brain).
Of course, it may be that all these
parameter-problems are just technical temporary problems. It may be that in future
even the {\it decoherence problem}, see, for example, [15], would be solved.
Nevertheless, there are doubts about
the possibility of the direct application of quantum {\it physical} theory to cognitive phenomena.

My personal critical attitude with respect to traditional quantum cognitive models is merely based on 
absence of a realistic bio-physical model that would
explain the transition from quantum processes in the microworld to cognitive processes. 
Of course, apart from the plethora of theoretical models such that proposed by Hameroff, Nanopoulos, 
Mavromatos, Mershin  and many others there are also very concrete experimental papers such as those
by Wolf  pertaining to exactly that point (see section 10 for the details). 
Nevertheless, at the moment we still do not have
a transition model that would be accepted by majority of neurophysiologists and psychologists.
\footnote{On one hand,
absence of such a realistic bio-physical model could be simply a consequence
of the huge complexity of this problem. If it is really the case, then we could expect that
in the future such a model would be finally created. On the other hand, it might occur that such a
model does not exist at all, since it might be that cognition is not generated by quantum 
processing of information in the microworld. As we have already mentioned,
the latter view has  become very popular over the last years. So absence of
a realistic transition model has  become a dangerous problem for the whole quantum cognition
enterprise. My personal position on the possibility of quantum reduction is based on  quantum as well
as neurophysiological experience. On one hand, I totally agree with neurophysiologists that there
are very strong arguments that cognitive information is prossesed on neuronal macro-level. On the
other hand, I am strongly impressed by the idea of interference of cognitive information proposed
by quantum physicists (e.g. Orlov, Penrose, Stapp, Hameroff and many others). I would like to explore
this idea and at the same time escape the problem of transition from quantum processes 
in the microworld to cognitive processes. I use the probabilistic calculus of quantum formalism (Hilbert
space formalism) to describe probabilistic amplitude thinking. In particular, in such a model, 
the brain would be able to create `interference of minds.' However, in my model elementary units of 
processing of cognitive information are macroscopic neuronal structures and not quantum micro systems.}

Finally, we discuss the holistic approach to cognitive phenomena based
on Bohmian-Hiley-Pylkk\"anen theory of active information. By considering
the pilot wave as a kind of information field they presented 
interesting models of cognitive processes, see [26]-[28], see also 
author's work [29]. In the latter paper there was proposed a mathematical model
of {\it field of consciousness}. This field is not defined on physical space-time. This
is a pure information structure. In principle, such a field can be considered as a mathematical
representation of Whitehead's {\it field of feeling} [1]. 

Consciousness-information models also  were developed
in books of M. Lockwood [30], and J. A. Barrett [31] 
(who use a many-minds version of many-worlds interpretation
of quantum mechanics) and in the author's paper [32] devoted to quantum information reality.

Over the last few years I tried to split, see [33]-[36], the quantum formalism into two more
or less independent parts: 

1) really {\it quantum} (quanta, Planck constant, discreteness);

2) Hilbert space {\it probabilistic formalism}. 

Pioneer investigations of M. Planck and A. Einstein on foundations of 
quantum theory (black body radiation and photoelectric effect) 
were merely investigations on discreteness (quantization)
of energy. Quantum probabilistic (Hilbert space) formalism was developed later (Born, Jordan, Heisenberg, 
Dirac [37]-[38]).
It was created to describe statistics of elementary particles. Due to such a historical origin, the
Hilbert space probabilistic calculus is always related to processes in the microworld.

However, careful analysis, [33]-[36], demonstrated that Hilbert space probabilistic calculus
(Born, Jordan, Heisenberg, Dirac, see e.g.  [37]-[38]) is 
a purely mathematical formalism that gives the possibility to work with context--dependent
probabilities, i.e., 
probabilities depending on complexes of physical conditions (contexts) related to concrete
measurements.
Therefore we could apply the Hilbert space probabilistic formalism, {\it quantum-like formalism,} not only to the 
description of statistical micro phenomena, but also to various phenomena outside
micro world. One of such possibilities is to apply quantum-like formalism to describe
statistical experiments with cognitive  systems. Here a quantum-like formalism describes probabilistic
distributions depending on neural, cognitive and social contexts.

Such an approach has no (at least direct) relation
to reductionist quantum models. We are not interested in statistical behaviour of micro 
 systems forming a macro system, the brain.  Therefore this approach does not
induce  such a problem as the transition from micro to macro (temperature, decoherence
and so on). We just use the Hilbert space  probabilistic formalism to describe cognitive measurements.
As in the ordinary quantum formalism, mental observables are realized as symmetric
operators in the Hilbert space of square integrable functions $\phi(q)$ depending on
the mental state $q$ of a cognitive system. By using the Hilbert space scalar product
we calculate averages of mental observables. Of course, this cognitive model is 
a purely statistical one. It could not provide a description of individual 
thought-trajectories.

One of the reasons for using quantum-like formalism to describe statistics
of measurements over cognitive systems is that cognitive systems (as well as quantum) are very sensitive
to changes of contexts of  experiments - complexes of physical and mental conditions ([33]-[36], compare
to Heisenberg [38] or Dirac [37]). Quantum-like formalism might be used to describe external measurements (in 
neurophysiology, psychology, cognitive and social sciences) over ensembles of cognitive systems or
neural ensembles in a single brain.\footnote{Thus we extend to cognitive sciences Heisenberg's viewpoint 
to the role of disturbances in producing of quantum interference.} As well 
as in quantum experiments with elementary particles, preparation of a 
statistical ensemble (of rats or people) plays the crucial role in cognitive measurements. Thus, as in 
ordinary quantum theory, it is meaningless to speak about a measurement without specifying 
a preparation procedure preceding this measurement. In cognitive sciences we also should follow 
Bohr's recommendation to take into account the whole experimental arrangement.
The main experimental evidence of quantum-like structure of statistical data obtained in 
neurophysiology, psychology, cognitive and social sciences should be interference of probabilities,
see [33]-[36] and section 10. 

Moreover, our quantum-like formalism can be used not only for describing external cognitive 
experiments, but also modeling of mentality. The basic assumption of our model is that the brain has the ability 
to ``feel" probabilistic amplitude $\phi (q)$ of information states produced by hierarchic neural pathways
in brain (and the whole body). There is also  a model of consciousness that
creates its context by performing self-measurements over extremely sensitive neural contexts.

One of the fundamental problems in foundations of cognitive quantum-like formalism is
the choice of a mathematical model for a mental configuration space on which a wave function 
is defined. We shall discuss this problem in the details in section 2. We now only remark
that the {\it Euclidean} physical space (within which the physical brain is located) does not look
attractive as a model of mental space.
Instead of this conventional model of space, we develop cognitive quantum-like formalism 
on the space of information strings that could be performed by chains of hierarchically ordered
neurons. Such a configuration space is geometrically represented by a hierarchic $p$-adic 
tree. In fact, this idea was already discussed in the authors's paper [32] (see also [39]-[44]). 
However, in [32] 
we did not use the standard Hilbert space formalism. We used a generalization of quantum probabilistic calculus
based on  $p$-adic probabilities. In the present paper we use the standard
Hilbert space formalism on $p$-adic trees. In fact, the mathematical formalism of $p$-adic
quantum mechanics is well developed, see Vladimirov, Volovich, Zelenov [45], [46],
see also [47]. We "simply" apply this formalism to 
cognitive phenomena.

In the ordinary quantum mechanics, we could go beyond 
the statistical application of quantum formalism.
One of the most attractive possibilities is to use the pilot wave Bohmian formalism.
As we have already remarked, the idea to use Bohmian mechanics in cognitive
sciences was already well discussed (Bohm-Hiley-Pilkk\"anen [26]-[28] and author [29]).
It is rather surprising that it seems to be impossible to create a variant
of the pilot wave extension of quantum-like mental formalism presented in this
paper. Formally we can introduce quantum-like mental potential and
force. However, there is no possibility to derive the equation of motion
(a kind of Newton equation) that would describe trajectories of individual 
mental states (describe "flows of mind").  In our formalism, this is 
a consequence of the mathematical structure of the model. However, it may be that 
there are some deep cognitive features behind  this mathematical result.

We start with some preliminary considerations on the choice of the geometry 
of a mental space.

\section{Where is consciousness located?}

The problem of location (or nonlocality) of consciousness
(as well as more primitive cognitive processes) is widely 
discussed in philosophic, neurophysiological and psychological 
literature, see
e.g. [48]-[56]. There is  large variety of views starting with such a primary
question: 

{\bf `Is consciousness located in human brain?'}

Both philosophic and neurophysiological discussions are, in fact,
related to one fixed geometry, namely the Euclidean one. 
It seems that such an approach was originated (at least in philosophy)
by Kant [55]. For him, the space was the absolute Euclidean space. He also pointed
out that the idea of space is the primary idea. Nothing could be even imagine
without any relation to space. Since Kantian space is identified with the Euclidean space,
we have to look for a place of consciousness in this space. It seems that 
this is the starting point of the main stream of modern philosophic, neurophysiological 
and psychological investigations. However, despite enormous efforts to find the place
of consciousness, there is more and more evidence that consciousness cannot be located in physical
space. What is wrong? I think the choice of geometry. I think that the use of the
Euclidean geometry is not adequate to this problem.

In fact, the idea that different natural phenomena are in general
described by using different geometries is well established in physics, especially 
general relativity and string theory. Following Chalmers [56],
we consider consciousness as a kind of natural phenomenon.
First we must find
an adequate model of a {\it mental space}. Then we  get the possibility
to describe cognitive (and conscious) phenomena. Let us imagine that we would like
to describe electromagnetic processes without using a mathematical model of the
electromagnetic field distributed on the Euclidean space. This seems to be impossible.
\footnote{Sometimes (especially in philosophy) 
one uses the expression 
`` to explain consciousness". I do not think that we can
``explain" it. In the same  way we can 
not ``explain", e. g., the electromagnetic field. 
We can
only describe mathematically and understand it 
via such a description, compare to Penrose [15], p. 419.}

We have already mentioned the use of various geometries in general in physics, e.g. in
general relativity and string theory.
However, these models are mainly locally-Euclidean (Euclidean manifolds).\footnote{Even the use of superspace
in superstring theory as well as in superfield theory cannot be considered as a fundamental
change of geometry. Locally, superspace is still a real continuous manifold, see e.g. [57] for the details.}
The use of such manifolds could not solve the 
problem of cognitive nonlocality (in particular, nonlocality of psychological
functions). One of possibilities is to proceed in a quantum-like way and 
use noncommutative mental coordinates, see B. Hiley [28]. Another possibility
is to try to find a model of classical mental configuration space
(probably as the basis of a quantum-like model). Since [39]-[46], we use
a purely information model of mental space, namely the space of all possible 
information strings that could be produced by hierarchically ordered
chains of neurons. One of the simplest models of such a space is a hierarchic
(homogeneous) $p$-adic tree ${\bf Z}_p,$ where $p$ is a natural number. It gives the number
of branches leaving each vertex of this tree. We remark that in mathematical models $p$
is typically a prime number, see [45], [47]. But this
is not so important for our cognitive considerations.

\section{Classical mental states produced by  one-layer brain}

{\bf 3.1. $p$-adic coding.}
We consider the simplest hierarchic ``brain" consisting of just one
hierarchic chain of neurons : 
$$
{\cal N} =(n_0, n_1,..., n_N,...).
$$ 
In a mathematical
model it is convenient to consider an infinite chain. 

In the simplest model
each neuron can perform only one of two states: $\alpha_j=1$ (firing) and
$\alpha_j=0$ (off). 

In more complex models each neuron $n_j$ can perform $p$ different
levels of activation: $\alpha_j=0,1,..., p-1.$ For example, such a coding can be obtained by using
frequencies of firing of neurons as basic elements of coding. Frequencies of firing 
are a better basis for the description of processing of information by neurons than
simple on/off. This has been shown to be the fundamental element of neuronal communication 
in a huge number of experimental neurophysiological studies (see e.g. [58], [59] on mathematical modeling
of brain functioning in the frequency domain approach).

One of possible $p$-adic coding models is the following one.
A $p$-adic structure associated with
frequency coding is generated in the following way. There exists some interval (of physical time) 
$\Delta$ (unit of ``mental time", see section 10 for further consideration). Then $\alpha_j$ is equal to
the number of oscillations of the neuron $n_j$ (in the hierarchic  chain ${\cal N})$ that 
are performed during the interval $\Delta.$ Here $p-1$ (where $p=p_{\Delta})$ is the maximally possible number
of oscillations during the period $\Delta$ that can be performed by neurons in the chain ${\cal N}.$
Thus in our model the $p$-adic structure of the brain 
of a cognitive system $\tau$ (that uses a frequency neural code)
is related to the time scale of the functioning of the brain, see also section 10.

We must mention one mathematical fact
that may be have some cognitive interpretation. The case $p=2$ is a very exceptional one
in $p$-adic analysis, see e.g. [45], [47].  We can speculate that the transition from 
2-adic coding  (firing/off) to more complex $p$-adic, $p> 3,$ coding (e.g. frequency
coding) was the evolutionary jump. Cognitive systems in the 
$p$-adic model exhibit essentially richer mental behaviour in the case $p> 2$ than 
in the case $p=2,$  see [47], [39]-[44] on classical mental dynamics.

{\bf 3.2. Hierarchy and ultrametricity.}  It is supposed
that neurons in a  layer ${\cal N}$ are hierarchically ordered: $n_0$ is the most important (igniting),
$n_1$  is less important and so on. The ${\cal N} $ is able to produce information
strings of the form:
$$
x=(x_0, x_1,..., x_N,...), \; x_j=0,1,..., p-1.
$$
We denote the set of all such strings by the symbol ${\bf Z}_p.$ The hierarchic structure
in the chain ${\cal N}$ induces a tree representation of ${\bf Z}_p.$ Information
strings are represented by branches of such a tree. 
\begin{center}
\begin{figure}[ht]
\unitlength1cm
\begin{picture}(12,5)
\put(2,3){\circle{0.5}}            \put(1.9,2.9){$\star$}
\put(2.5,2.8){\vector(2,-1){2.0}}
\put(2.5,3.2){\vector(2,1){2.0}}
\put(5,4.4){\circle{0.5}}            \put(4.9,4.28){$0$}
\put(5.5,4.35){\vector(4,-1){2.0}}
\put(5.5,4.45){\vector(4,1){2.0}}
\put(5,1.6){\circle{0.5}}            \put(4.9,1.48){$1$}
\put(5.5,1.55){\vector(4,-1){2.0}}
\put(5.5,1.65){\vector(4,1){2.0}}
\put(8,5.05){\circle{0.5}}          \put(7.9,4.95){$0$}
\put(8,3.7){\circle{0.5}}          \put(7.9,3.57){$1$}
\put(8,2.25){\circle{0.5}}          \put(7.9,2.15){$0$}
\put(8,0.9){\circle{0.5}}          \put(7.9,0.75){$1$}
\put(8.5,5.00){\vector(4,-1){1.0}}
\put(8.5,5.10){\vector(4,1){1.0}}
\put(8.5,3.65){\vector(4,-1){1.0}}
\put(8.5,3.75){\vector(4,1){1.0}}
\put(8.5,2.20){\vector(4,-1){1.0}}
\put(8.5,2.30){\vector(4,1){1.0}}
\put(8.5,0.85){\vector(4,-1){1.0}}
\put(8.5,0.95){\vector(4,1){1.0}}
\end{picture}
\caption{The $2$-adic tree}
\end{figure}
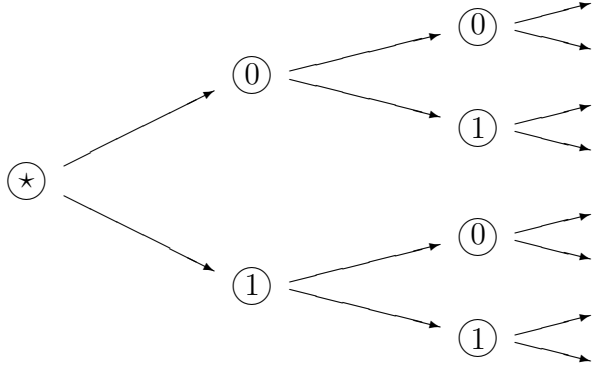
\end{center}
The distance between two branches,
$x$ and $y$, is defined in the following way. Let $l$ be the length of the {\it common root}
of these branches. Then the $p$-adic distance between $x$ and $y$ is defined as
$$
\rho_p(x,y)= \frac{1}{p^l}\;.
$$
Thus if $x=(x_j)$ and $y=(y_j)$ and $x_0=y_0,..., x_{l-1}= y_{l-1},$ but $x_{l}\not= y_{l},$
then $\rho_p(x,y)= \frac{1}{p^l}.$ This is a metric on the set of branches ${\bf Z}_p$ of
the $p$-adic tree. Two branches are close with respect to this metric if they have a sufficiently 
long common root. We remark that ${\bf Z}_p$ is complete with respect to the $p$-adic metric 
$\rho_p.$

The $p$-adic metric gives a topological representation of the hierarchic structure in 
neural chains. The distance between information strings $x$ and $y$ approaches the
maximal value  $\rho_p(x,y)=1$ if $x_0 \not = y_0.$ Thus the state (e.g. the frequency 
of firing) of the first neuron $n_0$ in a hierarchic  chain ${\cal N}$ plays the most important role.
States $x_j$ of neurons $n_j,$ where $j\to \infty,$ have a practically negligible contribution into the
geometry of the $p$-adic space.

The $p$-adic metric is a so called {\it ultrametric,} i.e., it satisfies the {\it strong triangle inequality}:
$$
\rho_p(x,y) \leq \max[\rho_p(x,z), \rho_p(z,y)],
x,y,z \in {\bf Z}_p.
$$
The strong triangle inequality can be stated
geometrically: {\it each side of a triangle is at most as long as the
longest one of the two other sides.} This property implies that all triangles are isosceles.
Ultrametricity is a very important feature of $p$-adic geometry. In fact,
ultrametricity is the exhibition of hierarchy. Recently it was proved in general topology
that in general case ultrametricity induces a treelike representation and vice versa, see
[60]. In many particular cases such a relation between ultrametricity and hierarchy was 
used in theory of spin glasses, see e.g. [61]-[63]. 

There exists a natural algebraic structure on this tree: addition,
subtraction and multiplication of branches. It is based on the representation
of information strings by so called $p$-adic numbers:
$$
x= x_0+ x_1 p +...+ x_N p^N+...
$$
This is the ring of $p$-adic integers. In particular, this is a compact
additive group. Thus there exists the Haar measure  $dx$ (an analogue of the ordinary
linear measure on the straight line). 

We set $B_r(a)=\{x \in {\bf Z}_p: \rho_p(x,a) \leq r\}$ and $S_r(a) =\{ x \in {\bf Z}_p: \rho_p(x,a)= r\},$
where $r=1/p^j,j=0,1,2,...$ and $a\in {\bf Z}_p.$ These are, respectively, balls and spheres in the metric
space ${\bf Z}_p.$ In particular, ${\bf Z}_p= B_1(0).$ Each ball has 
the structure of the homogeneous $p$-adic tree (scaling of the
basic tree given by ${\bf Z}_p).$

As in every ulrametric space, all these sets (balls and spheres) have a topological structure which seems
to be
rather strange from the point of view of our Euclidean intuition: they
are open and closed at the same time. Such sets are called {\it clopen.} 
Another
interesting property of $p$-adic balls is that  two balls have nonempty 
intersection iff one of these balls is contained in another. Finally we 
note that any point of the $p$-adic ball can be chosen as its center. 
Thus the ball is not characterized by its center and radius.

{\bf 3.3. Mental space.} We choose the space $Q = {\bf Z}_p$ as a {\it mental configuration space.}
Points $q \in Q$ are called mental 
classical-like states (or simply {\it mental states}) or mental positions.

Thus a mental state $q \in Q$ describes activity of neurons in a hierarchically ordered 
chain of neurons. This is a kind of information state. Such a state could not
be considered simply as the representation of physical (electro -- chemical) activity 
of neurons in a chain. There are two information parameters that play important roles in our model. 

First there is the hierarchic  structure in a neural chain. Neurons in a chain
``do not have equal rights." The igniting neuron $n_0$ is the bandmaster
of the orchestra ${\cal N}.$  This orchestra is rigidly hierarchic.
The next neuron $n_1$ in the  ${\cal N}$ is less important than $n_0$
and so on. I think that the presence of such a hierarchy plays an important role in 
creation of cognition and may be even consciousness. 

Another information parameter
is a natural number $p$ that determines the coding system of (one layer) brain 
${\cal N}.$ If we follow to the frequency approach to functioning of neural networks
in brain, then the parameter $p$ gives the maximal number of oscillation for 
a neuron in a chain ${\cal N}$ during the unit interval $\Delta$ of {\it mental 
time.} The $\Delta$ is an interval of physical time  that in our model determines
the neural code of ${\cal N},$ see section 10.

In our model a mental state provides only {\it cognitive representation} and 
not the {\it contents of consciousness.} All unconscious processes are performed on the 
level of mental states. We remark that in a multi-layer brain, see section 5.1, 
 various unconscious cognitive processes can be performed parallely. 
 Nonlinear dynamical models of 
such processes were studied in [47], [39]-[44]. One of the distinguishing features of $p$-adic
nonlinear dynamics is the absence of chaotic behaviour. In general $p$-adic dynamical systems are 
essentially more regular than real ones. Moreover, they are very stable 
with respect to random perturbations (in particular, noises), [43]. Typically a $p$-adic
random dynamical system has only deterministic attractors, see [43]. 
We remark that dynamics in spaces of $p$-adic numbers depends crucially on 
the parameter $p$ (determining the neural code of a brain  ${\cal N}).$
As it was discovered in [47],  the same dynamical system 
(e.g. given by a monomial $x^n)$  can demonstrate completely different behaviours for 
e.g. $p=2,$ or $p=3,$ or $p=1999,....$ 

We are now going to consider a quantum-like model based on the $p$-adic mental configuration
space $Q={\bf Z}_p.$ In particular, this model might be used to describe the transition
from unconscious representation of cognitive information to conscious one.

\section{Quantum-like formalism for one layer brain}

{\bf 4.1. Hilbert space probabilistic formalism for mental  observables.}
We consider the space of square integrable functions $L_2(Q, dx),$ where $Q={\bf Z}_p:$
$$
\phi: {\bf Z}_p  \to {\bf C}, \Vert \phi \Vert^2 = \int_{{\bf Z}_p} \vert \phi(x)\vert^2 d x < \infty.
$$
The space ${\cal H}=L_2(Q, dx)$ is chosen as the space of {\it mental quantum-like
states} (or {\it mental amplitudes}). These states are represented by normalized vectors $\phi \in {\cal H}:
\Vert \phi \Vert=1.$  The ${\cal H}$ is a complex Hilbert space with scalar
product 
\begin{equation}
\label{SK}
(\phi, \psi) =\int_Q \phi(x) \bar{\psi}(x) d x\;.
\end{equation}
{\it Mental observables} are realized as self-adjoint 
operators $A: {\cal H} \to {\cal H}.$ As in the ordinary quantum formalism,
by fixing a quantum-like state $\phi \in {\cal H}$ in general we do not fix the 
concrete value $A=\lambda$
of a mental observable $A.$ It is only possible  to find the average of $A$ in the state $\phi:$
\begin{equation}
\label{SK1}
<A>_\phi =\int_Q A(\phi)(x) \bar{\phi}(x) \;d x\;.
\end{equation}
However, if $\phi \in {\cal H}$  is an eigenfunction of $A$ corresponding to 
the eigenvalue $\lambda,$ i.e., $A \phi= \lambda \phi,$ then we can be sure
that we shall obtain the value $A=\lambda$ with probability 1.

The concrete representations of mental observables by self-adjoint operators
is a very important and nontrivial problem. This problem could not be solved by 
trivial generalization of an ordinary quantum formalism. We start with this surprising remark:
it seems to be {\it impossible to define mental position,} $q,$ observable. Formally the difficulty
is purely mathematical: we could not multiply a $p$-adic number $q \in Q$ with
a complex number $\phi(q).$ Therefore the standard Schr\"odinger's definition
of the position operator could not be generalized to the  cognitive case. Of course, we could try
to find some mathematical tricky ("non natural") definitions of mental position operator.
However, it might be that this mathematical difficulty is evidence of some important
feature of cognitive systems. It might be that: 

{\it Even in principle it is impossible
to measure mental states $Q$ of brain.}

In particular, we could not prepare  a statistical ensemble of brains having
the fixed mental state (there are no mental state eigenfunctions).

We can only find the probability that a mental state $q$ belongs to some 
(measurable) subset $O$ of the mental space $Q:$ $P(q \in O)= \int_O \vert \phi(x)\vert^2 d x.$

{\bf Example 4.1.} Let us consider the quantum like state $\phi \equiv 1$
(the uniform probability distribution of mental states). Then $P(q \in B_r(a))=r.$
Thus (as it could be expected) the probability to find this cognitive
system in the mental state $q$ belonging to a small ball around any 
fixed point $a$ is small.

{\bf 4.2. External mental measurements.} 
An important class of mental observables is given by measurements that are performed by external systems over
a cognitive system $\tau.$ In particular, in section 6 we shall introduce a neuron-activation 
observable that arises naturally in neurophysiological measurements. Besides neurophysiological
mental observables, we can consider e.g. psychological or social mental observables. In experiments
with people such a mental observable $A$ can be given just by a question $A.$ Here $A$ takes
two values: $A=1$, yes, and $A=0,$ no. In experiments  with animals values of $A$ give
possible reactions of animals to experimental conditions. In principle an external system that 
performs a measurement over a cognitive system $\tau$ need not be conscious nor even cognitive.
It can be, for example, a magnetic resonance device performing a measurement of neural activity.
\footnote{Such a viewpoint induces some difficulties, see, for example, 
Wigner [8] on the role of consciousness in quantum measurements.}

{\bf 4.3. Consciousness.}  I would not like to reduce mental measurements to external measurements. 
It is natural to try to describe consciousness as a continuous flow of mental self-measurements.
\footnote{``Continuous'' has the meaning of mental time continuity, see section 10.} The idea that 
cognitive representation of information in the brain becomes conscious in the process of self-measurements 
is not so new, see e.g. Orlov [13] for a quantum logic model of self-measuring consciousness:
{\small ``... the volitional act of a free choice plays in this theory a role analogous to the role
of the measurement act in quantum mechanics (with the important difference that the brain ``measures''
itself). Consciousness is a system which observes itself and evaluates itself -- being aware,
at the same time, of doing so.''}

The crucial point of our consideration is that we use a quantum-like ideology, instead of the
traditional quantum one. In our model the configuration space is the state space of 
macroscopic neural networks. Thus we need not go deeply into the microworld to find the origin of 
consciousness (e.g. no collapses of mass-superpositions and so on). So we need not apply to
quantum gravity (or even superstring theory). 

In our model it is supposed that each cognitive  system $\tau$ developed the ability to feel 
the probability distribution $P(q)$ of realization of the hierarchic
information string $q$ by its neural system. Such an ability is basically transferred from generation to
generation. However, for each $\tau$ it is permanently developed in the process of brain's functioning.
This probability distribution $P(q)$ has an amplitude $\phi(q)$ that can be mathematically described by 
a normalized vector in the Hilbert space ${\cal H}=L_2(Q, dx).$ As usual, $P(q)= \vert \phi(q)\vert^2.$ 
As was already discussed, the appearance of the quantum-like probabilistic formalism
(instead of classical Kolmogorov probabilistic formalism) is a general consequence of sensibility
of $P(q)$ to changes in the neural context. Here
$$
\phi(q)= \sqrt{P(q)} e^{i\theta(q)} .
$$
Here $\theta(q)$ is a phase parameter. It appears automatically in transformation
of probabilities from one mental representation (see 5.4) to another, see [33]-[36].
We shall illustrate the role of $\theta(q)$ on the example of transition from mental-position
to motivation representation, section 5.4.

In our model ``feeling" of the probability distribution is performed  on an unconscious level.
In particular a cognitive system does not feel consciously the evolution of the mental amplitude
$\phi(t,q).$

Moreover, we suppose that each conscious cognitive system $\tau$ has the ability to perform self-measurements.
Results of these measurements form the contexts of consciousness. I do not try 
to develop such a model of consciousness in the present paper. The main aim of this
paper was to  present quantum-like formalism corresponding to {\it hierarchic} neural networks.
In principle, the reader can use only a restricted viewpoint of mental observables 
as external measurements over cognitive systems. We just consider a possible scheme
of functioning of such a (quantum-like)  self-measuring consciousness. 

{\bf 4.4. Random dynamical quantum-like consciousness.} Let us denote the set of all operators 
representing mental observables participating in the creation of the contents of 
consciousness by the symbol ${\cal L}_{\rm{cons}}({\cal H}).$ Let us consider 
a random dynamical system (RDS: see, for example, [64] for  general theory) that at each
instant of (mental) time  chooses randomly some set of commutative operators
$A_1,..., A_m \in {\cal L}_{\rm{cons}}({\cal H}).$ The contents of consciousness
at this instant of time is created by the simultaneous measurement of $A_1,..., A_m.$

One of the main distinguishing features of  the RDS-model is that a  RDS
in the space of mental observables can have long range 
memory. Such a feature of RDS is very important to create a realistic 
mathematical model of functioning of consciousness. Our consciousness does not
consist of discrete moments but there is flow of consciousness. We remember something about
our earlier conscious experiences, see e.g Whitehead's analysis of this problem [1], p. 342-343:
{\small ``Whenever there is consciousness there is some element of recollection. It recalls earlier phases from the dim
recesses of the unconscious. Long ago this truth was asserted in Plato doctrine of reminiscence. 
No doubt Plato was directly thinking of glimpses of eternal truths lingering in a soul derivative from timeless 
heaven of pure form.
Be that as it may, then in a wider sense consciousness enlightens experience which precedes it, and could be without
it if considered as a mere datum. Hume, with opposite limitations to his meaning, asserts the same doctrine....But the immediate point is the
deep-seated alliance of consciousness with recollection both for Plato and fo Hume."}

\section{Motivation observable}

{\bf 5.1. Multi-layers hierarchic brain.} 
To consider nontrivial examples of mental observables,
it is convenient to study a "brain" having a more complex 
mental space. Such a brain  consists of a few hierarchic $p$-adic trees.
We consider a layer of neurons 
$$
{\cal N}= (..., n_{k},..., n_0, ..., n_l,...)
$$
that goes in both directions (in the mathematical model it is infinite
in both directions). Each neuron $n_j, j=0,\pm 1, \pm 2,...,$ can be 
the igniting neuron for right hand side hierarchic chain: ${\cal N}_j=(n_j,...., n_l,...).$
The corresponding mental space ${\bf Z}^{(j)}$ consists of all information strings
$$
x=(x_j,x_{j+1}, ..., x_l,...), x_l=0,1,..., p-1
$$
(in particular, ${\bf Z}_p= {\bf Z}^{(0)}).$ Each  space has the structure
of the homogeneous $p$-adic tree. These spaces are ordered by inclusion: ${\bf Z}^{(j+1)}\subset {\bf Z}^{(j)}.$
We consider union of all these space ${\bf Q}_p= \cup_{j=-\infty}^\infty {\bf Z}^{(j)}.$ 
Geometrically this space is represented as a huge collection of trees ordered by the inclusion
relation.
On this space we
can introduce the structure of ring: addition, subtraction and multiplication of branches of trees.
If the coding parameter $p$ is a prime number (i.e., $p=2, 5, 7,..., 1997, 1999,...),$ then
${\bf Q}_p$ is a field, i.e., division of branches also is  well defined. In this case
${\bf Q}_p$ is a number field (of $p$-adic numbers). Arithmetical operations are performed
by using $p$-adic number representation of branches:
\begin{equation}
\label{S}
x= \sum_{i=j}^\infty x_i p^i, j=0, \pm 1, \pm 2,...
\end{equation}
Metric on ${\bf Q}_p$ is defined in the same 
way as on ${\bf Z}_p.$ In particular, each tree ${\bf Z}^{(j)}$ 
coincides with a $p$-adic ball $B_r(0),$ where $r=1/p^j.$
We shall also use $p$-adic absolute value: $ \vert x \vert_p= \rho_p(x, 0).$ 
To calculate it, we have to find in the chain ${\cal N}$ the first (from the left hand side) firing neuron $n_j$
($x_j \not= 0,$ but $x_l=0$ for all $l < j)$ and set  $\vert x\vert_p = 1/p^{j}.$

The ${\bf Q}_p$ is a locally compact field. Hence, there also exists the Haar measure $d x.$

We now choose $ Q ={\bf Q}_p$ as a model of a mental configuration space; consider the Hilbert ${\cal H}=
L_2(Q, d x)$ of square integrable functions $\phi: Q\to C$ as the space of quantum-like mental states.

{\bf 5.2. Motivation magnitude  observable.} It would be interesting to consider the following 
quantity (more precisely, qualia): {\it motivation} $\xi$ to change the mental state $q.$ Unfortunately,
for the same reasons as for the mental state observable we could not introduce 
a motivation observable. However, we can introduce an observable $M_\xi$ that will give
the magnitude of a motivation. It is impossible to prepare a brain with the fixed
motivation $\xi$, but we could prepare a brain with the fixed amplitude of a motivation (that gives 
a measure of motivation's strength). Such $M_\xi$ must be a kind of derivative with respect to the mental state
(coordinate) $q.$ Such a generalization of derivative is given by 
{\it Vladimirov's operator} $D$, see [45], defined with the aid of the $p$-adic 
Fourier transform.\footnote{We remark that it is
impossible to define  the derivative for maps from ${\bf Q}_p$ to ${\bf R},$
see [47].}

{\bf $p$-adic Fourier transform:}
$$
\tilde{\phi}(\xi) = \int_{Q} \phi(x) e(\xi x) d x, \xi \in Q,
$$ 
where $e$ is a $p$-adic character (an analogue of exponent): 
$e(\xi x)= e^{2\pi i \{ \xi x\}}.$ Here, for a $p$-adic number
$a,$  $\{a\}$ denotes its fractional part, i.e., for 
$a= \frac{a_{-m}}{p^m}+...+\frac{a_{-1}}{p} + a_0 +...+ a_k p^k+...$
(where $a_j=0,1,..., p-1,$ and  $a_{-m}\not= 0)$ we have
$$
\{a\}=\frac{a_{-m}}{p^m}+...+\frac{a_{-1}}{p}.
$$

{\it Vladimirov's operator} of order $\alpha > 0$ is defined as 
$$
D^\alpha(\phi)(x) = \int_{Q} \vert \xi \vert^\alpha \tilde{\phi}(\xi) e(-\xi x) d \xi.
$$
We remark that $D^\alpha D^\beta = D^{\alpha+\beta}.$  We define the motivation magnitude observable $M$ as

$M_\xi= h D$

Here $h= \frac{1}{p^m}$ is some normalization constant.
The $h$ plays the role of the Planck constant in
ordinary quantum mechanics. At the moment it is not clear: ``Can we expect that there exists 
a kind of universal constant $h$, the {\it mental Planck constant}?" I am quite sceptical that 
such a universal normalization constant really exists. It is more natural to suppose 
that $h$ would depend on a class of cognitive systems under consideration. In fact,
by finding $h$ (the level of motivation discretization) we find the basis $p$
of the coding system.

To calculate averages of the momentum magnitude operator $M_\xi$ for different
quantum-like mental states, it is natural to use the Fourier transform.
By analogy with ordinary quantum mechanics we 
could say: to move from position to momentum representation.

{\bf Example 5.1.} Let a quantum-like state $\phi$ is such that its Fourier 
transform $\tilde{\phi}(\xi)$ is uniformly distributed over the ball
$B_r(0), r=1/p^l.$ Here
$$
<M_\xi>_\phi= p^l \int_{B_r(0)} \vert \xi\vert_p d \xi=
\frac{1}{p^{l-1}(p+1)}.
$$

{\bf 5.3. Wholeness of mental observables.}
It is important to remark  that (in the opposite to the ordinary 
quantum momentum) the $M_\xi$ is {\it nonlocal operator.} It can be represented as an integral
operator, see [45]:
$$
D(\phi)(x)=\frac{p^2}{p+1} \int_Q \frac{\phi(x)- \phi(y)}{\vert x - y \vert_p^2} d y\;.
$$
To find $M_\xi(\phi)(x)$ in some fixed point $x,$ we have to take into account
values of $\phi$ in all points of the mental configuration space. Thus {\it motivation psychological
function} can not be localized in some particular neural substructure of brain.
 
This example is a good illustration of the mathematical description of nonlocality of psychological functions
in our $p$-adic quantum-like model. One of the main distinguishing features of this model is nonlocality
of derivation operator (Vladimirov's operator). Hence  the corresponding psychological
function is produced by the whole
neural system of the body (as indivisible system).

{\bf 5.4. Psychological functions as quantum-like representations.} The mathematical description of the 
motivation psychological function by using a new representation in the Hilbert state space
is the basic example that can be generalized  to describe all possible psychological functions. We remark 
that the motivation representation is, in fact,  a new system
of quantum-like mental coordinates. In the case of motivation a new system of coordinates was generated
by a unitary operator in the Hilbert state space, namely Fourier transform. 

In  the general case each psychological function $F$ is represented mathematically by choosing a
system of coordinates in the Hilbert state space, {\it mental representation.} 
Thus we can identify the set of all
psychological functions with the set of all unitary operators: $U\to F_U$ and $F\to U_F.$
All mental observables $A$ represented by 
self-adjoint operators  that can be diagonalized by using the concrete $U$-representation
can be related to the corresponding psychological function $F_U.$ For example, concerning
visual function observables of shape and colour can be diagonalized in the visual representation
of the state Hilbert space.

In such a model all psychological functions coexist peacefully in the neural system. 
The evolution of the quantum-like mental state $\phi(t,x)$ (see section 9) induces 
the simultaneous evolutions of all mental functions (in this state). This is unconscious evolution.
Thus a conscious system $\tau$ are not consciously aware about simultaneous evolution of the various psychological functions.
Only by the acts of self-measurements of some mental observables $A_F^{(j)}$ that are
diagonal in the $F$-representation the $\tau$ becomes aware about some features of the corresponding psychological
function $F.$

 {\bf 5.5. Free mental waves.} We remark that Vladimirov's operator $D$ has a system of (generalized) eigenfunctions that is similar
 to the system of free-wave eigenfunctions in ordinary quantum mechanics, where 
 $\phi_\xi(x)=e^{i \xi x/h}$ corresponds to the fixed value $\xi$ of momentum.  In the mental framework:
 
 $M_\xi e(h \xi x)= \vert \xi \vert_p  e(h \xi x).$
 
 Here we have used the fact [45]: $D e(\xi x)= \vert \xi \vert_p  e(\xi x).$
 We remark that in the ordinary quantum formalism the $h$ is placed in denominator,
 $\xi x/h,$ and in the $p$-adic quantum formalism it is placed in the nominator,
 $h \xi x.$ This is a consequence of the fact that $1/h$ is large in ${\bf R}$ 
 and $h$ is large in ${\bf Q}_p.$

 The function $\phi_\xi(x)=e(h \xi x)$  is a kind of {\it free mental wave}
 corresponding to the fixed value $\xi$ of the motivation. As $\vert \phi_\xi(x) \vert=1$
 for all $x\in Q,$ the probability to find a cognitive system in the mental state
 $x$ does not depend on $x.$ By analogy with the ordinary quantum mechanics we would like to
 interpret this mathematical fact in the following way:
 By fixing the magnitude of motivation (strength of willing)
 we could not localize the mental state. However, we will 
 see soon that such an analogy (between material and mental
 states) can not be used.
 
 A free mental wave $\phi_\xi$ gives a good example illustrating of the role of the phase $\theta(x)$
 of the mental amplitude. Here we have 
 \begin{equation}
 \label{PH}
 \theta(x)= 2\pi \{ \xi x\} .
 \end{equation}
 Thus if a cognitive system $\tau$ has the fixed motivation $\xi$ and the mental probability
 distribution $P(x)$ is uniform, then the phase of the corresponding mental amplitude 
 is determined by (\ref{PH}). Thus in general the phase $\theta(x)$ of a mental amplitude
 $\phi(x)$ is not the pure product of neural  activity. This phase contains information on the 
 transition from one mental representation to another. 
 
 {\bf 5.6. Privacy of motivation states.} The wave $\phi_\xi(x)$
 is not determined uniquely by the observable $M_\xi.$ The main distinguishing feature
 of $p$-adic quantum mechanics (discovered by Vladimirov, [45]) is 
 the huge degeneration of the spectrum of the momentum and energy operators. 
 In particular, beside eigenfunctions $\phi_\xi(x),$ the $M_\xi$ has  an infinite
 set of other eigenfunctions corresponding to the eigenvalue
 $\lambda= \vert \xi \vert_p (= p^k$ for some $k=0, \pm 1,\pm 2,...).$ 
 
 Each $\lambda= p^k, k=0,\pm 1, \pm 2,...$ corresponds to an infinite series of 
 eigenfunctions (distinct from the free mental wave $\phi_\xi(x))$ 
 belonging to $L_2(Q, d x).$\footnote{We remark that
 free mental waves $\phi_\xi(x)$ are so called generalized eigenfunctions.
 They are not square integrable.
 Thus they do not belong to the space of quantum-like mental states ${\cal H}=
 L_2(Q, d x).$ One could speculate 
 that such non-normalizable free mental waves may be related to altered consciousness events such as e.g. 
 hallucinations.} These  eigenfunctions are well localized (concentrated in balls) in 
 the mental configuration space. 
 
 This is very natural from the mental point of view. It would be quite strange if the 
 only quantum-like mental state with the fixed motivation magnitude is the state
 $\phi_\xi$ characterized by totally indefinite distribution of mental states $q.$
 By intuitive reasons there must be quantum-like mental states characterized 
 by the fixed $M_\xi= \lambda$ that are concentrated on a special class of mental states (a kind 
 of special mental activity).
 
 One of the most important distinguishing features of quantum-like mental theory is that 
 the motivation magnitude operator $M_\xi$ has a discrete spectrum (except of
 one point, see further considerations).
 Hence the magnitude of the motivation does not change continuously.
 
 There exists only one point of the spectrum of the operator $M_\xi$ that is not its eigenvalue:
 $\lambda=0.$ It is the limit point of the eigenvalues $\lambda_k= p^k, k \to \infty.$ There is no
 eigenfunction $\phi_0$ belonging to the state space ${\cal H}.$  Thus our model brain
 could not be (alive, awake?) in the stationary quantum-like mental state having the motivation of 
 zero magnitude.

 Another distinguishing feature is infinite degeneration of spectrum. This purely mathematical
 result can have important implications for the problem of {\it correspondence between mental and physical
 worlds.} In fact, due to this huge degeneration, {\it  we could not uniquely determine 
 the mental state of a cognitive system by fixing the  motivation magnitude $M_\xi.$}

  \section{Neuron-activation observable}
  
  As we have already discussed, we could not introduce a mental state observable $q.$
  However, in the same way as for the motivation we can introduce an operator of the $p$-adic
  magnitude of a mental state:
  $$
  M_q \phi(x) = \vert x \vert_p \phi(x).
  $$
  Spectral properties of this operator are similar to spectral properties of the operator $M_\xi:$
  discreteness and infinite degeneration of spectrum. Eigenfunctions of $M_q$ (belonging to 
  ${\cal H}= L_2(Q, d x))$ are localized in $p$-adic balls--trees. Therefore:
  
  {\it There exist stationary states of $M_q$ that are characterized by activation of the fixed tree
  of mental states.}
  
  Unfortunately, $M_q$ could not be used to fix such a tree (as a consequence of infinite
  degeneration of spectrum).
  
  The operators of position and motivation magnitudes, $M_\xi$ and $M_q,$ 
  do not commute (as operators of position and momentum in ordinary
  quantum mechanics):
  $$
  [M_q, M_\xi]= M_q M_\xi -  M_\xi M_q = h J,
  $$
  where $J\not =0$ is an integral operator [45]. Thus we get a {\it mental uncertainty
  relation}, compare to [32]:
  
  {\it For any quantum-like mental state $\phi,$ 
  it is impossible to measure motivation and position magnitudes with an arbitrary
  precision.}
  
  By measuring the motivation magnitudes we change position magnitudes and vice versa.
  This can also be   expressed mathematically by using the $p$-adic Fourier transform.
  We denote by $\Omega_r(x)$ the characteristic function of the ball $B_r(0)$
  (it equals to 1 on the ball and 0 outside the ball). We have [45], p. 102,
  $\tilde{\Omega}_r(\xi)= \frac{1}{r} \Omega_{\frac{1}{r}}(\xi).$
  If the state of mind is concentrated on the ball-tree $B_r(0),$ then motivations are concentrated on
  the ball-tree $B_{\frac{1}{r}}(0).$
  
 As in the case of the $M_\xi$-observable, the point $\lambda=0$ belongs to a non discrete spectrum of 
 the $M_q$ observable. Thus there is no stationary quantum-like mental  state $\phi$ corresponding
 to zero magnitude of $q.$ A cognitive system is not alive (awake?) in such a state.
 
 To understand better the mental meaning of the $M_q$-observable, it is useful to consider
 a new mental observable:
 
 $A = - \log_p M_q.$
 
 If, $\phi\in {\cal H}$ is an eigenstate of the $M_q$ corresponding to the eigenvalue $\lambda= \vert q\vert_p =
 \frac{1}{p^k},$ then $\phi$ also is an eigenstate of $A$ corresponding to the eigenvalue
 $\mu = k$  and vice versa. Thus the discrete part of the $A$-spectrum coincides with the set of integers
 ${\bf Z}.$ The $A$ gives the position of the igniting neuron in a layer of neurons. It is called
 {\it neuron-activation observable.} We note that there is an interesting relation between neuron-activation observable and entropy.
 
 Let us consider the quantum-like state $\phi(q)= \sqrt{(p+1) \vert q\vert_p} \Omega_1(q).$
 Here $\sqrt{p+1}$ is just the normalization constant. The corresponding probability distribution
 ${\bf P}(q)= (p+1) \vert q\vert_p$ on the tree ${\bf Z}_p$ and equals to zero outside this tree.
 The entropy of this probability distribution 
 $$
 E_{{\bf P}}= - \int_{{\bf Z}_p} \log_p {\bf P}(q) {\bf P}(q) d q = <A>_\phi - \log_p(p+1).
 $$
 
 \section{Complex cognitive systems; evolution}
 
 We now consider a cognitive system consisting of $n$ hierarchic layers of neurons.
 It can be an individual brain as well as a system of 
 brains. The mental space of this cognitive system is 
 $$
 Q= {\bf Q}_p \times \cdots {\bf Q}_p
 $$
 ($n$ times). For each mental coordinate $q_j, j=1,2,...,n,$ we introduce
 the motivation magnitude operator $M_j = h D_j,$ where $D_j$ is Vladimirov 
 operator for $q_j.$ 
 We introduce {\it kinetic mental energy} (free energy of motivations) as
 
 $H = h^2 \Delta,$ 
 
 where $\Delta=\sum_{j=1}^n D_j^2$ is {\it Vladimirovian}
 (a $p$-adic analogue of the Laplacian).
 
 We note that free mental waves $\phi_\xi(x)=e(h \xi x)$ are eigenfunctions of 
  this operator with eigenvalues $\lambda= \vert \xi \vert_p^2.$ As in the cases of 
 the $M_q$, $M_\xi$-observables, there is an infinite family of other eigenfunctions distinct
 from free mental waves. These functions are localized on the mental configuration space
 (describing fixed ranges of ideas).  The spectrum is discrete : $\lambda= p^k, k=0,\pm 1, \pm 2.$ 
 Thus the kinetic mental energy is changed only by jumps. 
 The $\lambda=0$ is the only point that belongs to the non discrete spectrum of the  operator of the kinetic
 mental energy.

 Interactions between brain's layers as well as interactions with the
 external world are described by the  operator of the potential mental energy.
 It is given by a real valued function (potential) $V(q_1,...,q_n).$ The total mental
 energy is represented by the operator:
 
 $H= h^2 \Delta + V.$
 
 We note that a mental potential $V(q_1,...,q_n)$ can change crucially spectral
 properties of the mental energy observable. If $V$ depends only on $p$-adic magnitudes
 $\vert q_j \vert_p$ of mental coordinates and 
 $V\to \infty ,\vert q_j \vert_p \to \infty,$ and $V$ is bounded from below (e.g. nonnegative),
 then the spectrum of $H$ (that is discrete) has only finite degeneration.
 Thus the "state of mind" of a free cognitive system could not be determined by fixing 
 the mental energy. However, by using additional mental (information)
 potentials we could (at least in principle) do this.
 
 The ground mental energy
 state $\lambda_0$ is not degenerated at all. In the latter case by fixing the minimal value
 of the mental energy $H= \lambda_0$ we can determine the "state of mind", namely 
 the $\lambda_0$-eigenstate. Even for other eigenvalues we can try to determine 
 the "state of mind" if the degeneration of spectrum is not so large. 
 It is interesting to remark that mathematical results [45] imply that degeneration 
 of eigenvalues (distinct from the ground energy) increases (as $p^2$) with increasing of $p.$
 If we connect the complexity of a cognitive system with the coding base $p,$
 then we obtain that, for complex cognitive systems (e.g. $p=1999),$ 
 it is practically impossible to determine the "state of mind" corresponding to the fixed value of mental energy.

 \section{Entanglement of psychological functions}
 
 {\bf 8.1. Classical viewpoint to localization of psychological functions.} 
 The problem of neural localization of  psychological functions split the neurophysiological community,
 see e.g. A. R. Damasio [53]:{\small ``One held that psychological functions such as language or memory could never be traced 
 to a particular region of brain.
 If one had to accept, reluctantly, that the brain did produce the mind, it did so as a whole and not as a collection 
 of parts with special functions. The other camp held that, on the contrary, the brain did have specialized parts and
 those pars generate separate mind functions."} Both adherents of wholeness and localization of psychological functions 
 have a lot of experimental evidences supporting their views. A kind of peaceful unification of these two views
 to localization of psychological functions is given by our model of coding of cognitive information by hierarchic
 pathways activity. As each pathway ${\cal N}$ is hierarchic, there are a few neurons in the pathway
 that play the most important
 role. Their location in some domain $U$ of the brain determines localization of a psychological 
 function containing ${\cal N}.$ However,
 ${\cal N}$ goes throughout many other brain (and body) regions. So $U$-localization is only a
 kind of {\it fuzzy localization.}
 
 However, I do not think that this is the end of the localization story. We suppose that cognition involves 
 not only classical dynamics of neural networks, but also quantum-like processing described by the evolution of 
 quantum-like wave function, see section 9. The latter gives the amplitude of probability distribution
 of realization of classical mental states (hierarchic strings of e.g. frequencies of firings).
 Such a quantum-like processing of cognitive information would automatically create psychological functions
 that do not have even fuzzy localization. Such functions are induced via entanglement of localized 
 psychological functions.
 
 {\bf 8.2. Entanglement.} Let $U_1, ..., U_k$ be some neural structures -- ensembles of hierarchic
 neural pathways -- specialized for performing  psychological functions $F_1,..., F_k.$ Consider corresponding 
 Hilbert spaces of quantum-like mental states: ${\cal H}_j= L_2(Q_j, d x_j),$ where $Q_j={\bf Q_p}^{n_j}$ and
 $ d x_j$ is the Haar measure on $ Q_j.$ Let $e_{F_j}$ be the orthonormal basis in ${\cal H}_j$
 corresponding to the function $F_j.$ 
 
 Let us consider the Hilbert space of quantum-like mental states of the composite neural system, 
 $U= U_1 \cup ...\cup U_k:\; {\cal H}=  L_2(Q, d x),$ where $Q= Q_1 \times... \times Q_k.$ Here  a normalized
 state $\phi(q_1,..., q_k)$ gives the amplitude of probability that $U_1$ produces $q_1$,...,
 $U_k$ produces $q_k.$  We consider in ${\cal H}$ the orthonormal basis $e$ obtained as the
 tensor product of bases $e_{F_j}.$ The $e$ describes a mental representation corresponding to the psychological function
 $F=(F_1,..., F_k)$ produced by classical combination of  psychological functions $F_1,..., F_k.$
 Let us now consider
 some other basis $\tilde{e}$ containing nontrivial linear combinations of vectors of the $e.$ 
 This basis gives the mental representation of a psychological function $G$ that could not be reduced
 to classical combination of psychological functions $F_j.$ We call $G$ an {\it entanglement of 
 psychological functions} $F_j.$ Of course, $G$ is produced by the collection $U$ of
 neural structures $U_j.$ But $G$ arises as nontrivial quantum-like combination of psychological functions
 $F_j.$ 
 
 We remark that entanglement of psychological functions has nothing to do with entanglement of quantum states of
 individual micro systems in the brain (compare to conventional reductionist quantum models of brain functioning).
 Entanglement of psychological functions is entanglement of probabilistic amplitudes for information states
 of macroscopic neural systems, see section 10 for further discussion.

 \section{State-evolution}
 
 We want to describe the evolution of a quantum-like mental state
 (mental wave function) $\phi(t,x).$ The first natural and rather 
 nontrivial problem is the choice of the evolution parameter $t.$ 
 This problem was discussed in detail in [32]. It was shown that
 there are different natural possibilities to describe the evolution of 
 mental states: "mental time", "psychological time" as well as ordinary
 physical time evolution, see also section 10. In this paper we consider the evolution
 with respect to physical time $t$ belonging to the real line 
 ${\bf R}.$ To derive the evolutional equation for $\phi(t,x),$ 
 we proceed in the same way as Schr\"odinger in ordinary quantum mechanics.
 We start with a free mental wave 
 $\phi_\xi(x)= e(h \xi x), \xi, x \in {\bf Q_p}.$ We have:
 
 $H_0 \phi_\xi(x)= \vert \xi \vert_p^2 \phi_\xi(x),$ where $H_0= h^2 D^2$
 is the operator of the mental energy for a free system.
 
 The $\phi_\xi(x)$ is a stationary state corresponding to mental energy
 $E= \vert \xi\vert_p^2.$ Such a wave evolves as 
 
 $\phi_\xi(t, x) = e^{\frac{iEt}{h}} \phi_\xi(x).$
 
 We note that this function is a combination of two essentially
 different exponents: ordinary exponent and $p$-adic character. This function 
 satisfies to the evolutional equation:
 \begin{equation}
 \label{SH}
 i h \frac{\partial \phi}{ \partial t}(t,x) = h^2 D^2 \phi(t,x).
 \end{equation}
 This is {\it Schr\"odinger's mental equation} for a free cognitive system.
 If we introduce a mental potential $V(x)$, then we get general  Schr\"odinger's mental equation:
 \begin{equation}
 \label{SH1}
 i h \frac{\partial \phi}{ \partial t}(t,x) = h^2 D^2 \phi(t,x) + V(x) \phi(t,x).
 \end{equation}
 If the initial quantum-like state $\psi(x)= \phi(0,x)$ is known, then
 by using (\ref{SH1}) we can find $\phi(t,x)$ at each instant $t$ of physical time.
 Under quite general conditions [45], the operator $H=  h^2 D^2 + V(x) $
 is a self-adjoint operator. Therefore (\ref{SH1}) is the standard Schr\"odinger's 
 equation in the Hilbert space ${\cal H}$ for one rather special class of operators
 $H.$ There also are mathematical results on analytical properties
 of solutions and correctness of Cauchy problem [47].
 
 {\bf Remark 9.1.} (Bohmian theory) {\small We can try to develop an analogue 
 of Bohmian (pilot wave) approach. As in ordinary Bohmian mechanics,
 we can define a quantum-like mental potential 
 \begin{equation}
 \label{SH2}
 W_ \phi(t,x) = -\frac{h^2}{R} D^2 R, \mbox{where}\; R(t,x) =\vert \phi(t,x) \vert .
 \end{equation}
 This potential has the same properties as the ordinary quantum potential:
 (a) $W_\phi(t,x)$ does not depend on the absolute magnitude of $\phi$;
 (b) $W_\phi(t,x)$ depends on the second variation of the magnitude of $\phi.$ 
 However, (in the opposite to ordinary Bohmian mechanics) we could not describe evolution of an individual mental 
 state (position) $q(t)$  by using Newton's equation with additional potential $W.$ 
 At first glance,
 this is a purely mathematical difficulty. But I think that this mathematical fact has deep cognitive
 meaning, namely that the dynamics of quantum-like state $\phi(t,x)$ does not determine the dynamics of classical
 mental states. Very different flows of classical mental states (hierarchically ordered neural flows) can produce
 the same wave $\phi(t,x).$ In our model only this wave determines results of mental measurements.
 Thus (in our model) it seems to be impossible to find a one to one correspondence between mental behaviour
 and neural activity. The flow of  consciousness does not uniquely correspond to neural dynamics in the brain.}

\section{Discussion}

{\bf 10.1. Why quantum-like formalism?}
One of the main reasons to expect that mental observables (including mental self-observables)
should be described by the quantum-like (Hilbert space probabilistic) formalism is the very high
sensivity of neural structures to changes of contexts of measurement. Such a sensivity
implies the violation of rules of classical probabilistic calculus and induces a so called 
quantum probabilistic calculus, see [33], [35] for the detailed analysis. The main distinguishing feature
of this quantum probabilistic calculus is interference of probabilities of alternatives. Therefore
a quantum-like structure of mental observables should imply interference effects for such observables.
In [65] a  general scheme of mental measurements was proposed that could be used to find the interference
effect. It may be that the corresponding statistical data have already been collected somewhere. We need
only to extract the interference effect. 

Another reason for quantum-like considerations is the discrete structure of information processing in brain.
It is natural to describe this exchange by quanta of information by a quantum-like formalism. In particular,
in our model we automatically obtained that basic mental observables such as e.g. mental energy
have discrete spectra. We underline that {\it philosophy of organism} of Alfred Whitehead was one of the first
philosophic doctrines in that fundamental proto-mental elements of reality, namely 
{\it actual occations,} had quantum (in the sense of discreteness) structure. The philosophy of organism 
was based on one-substance cosmology, see [1], p. 26, {\small  ``Descartes and Locke maintained a two-substance ontology --
Descartes explicitly, Locke by implication. Descartes, the mathematical physicists, emphasized his account to of corporeal substance; 
and Locke, the physician and the sociologist, confined himself to an account of mental substance. The philosophy of 
organism, in its scheme for one type of actual entities, adopts the view that Locke's account of mental 
substance embodies, in a very special form, a more penetrating philosophic description than Descartes' account of
corporeal substance."}

{\bf 10.2. Quantum-like statistical behaviour  and consciousness.}
In our model of consciousness  as the process of (quantum-like) self-measurements over hierarchic
neural structures the quantum structure plays an important, but not determining role. There are 
many sensitive physical systems (not only microscopic, but also macroscopic) that could exhibit quantum-like
behaviour, see [33]-[35] for the details. Thus to be quantum-like 
is not the sufficient condition to be conscious.
There must be something else that is crucial in inducing consciousness. This consciousness determining factor
may be quantum as well as classical (or a very special combination of classical and quantum factors). 

It seems that the crucial point might be the ability to ``feel" the ensemble probability distribution 
of information strings produced by neural activity. My conjecture is that such a feeling is the basis
of mentality. In such a model a cognitive system reacts not to  firings of individual neurons or even
large populations of neurons, but to {\bf integral probability distributions of firings.} If this 
is the really the case, then quantum-like probabilistic formalism would appear automatically, since
this is the most general theory of  transformations of context depending probabilities [35]-[39].

{\bf 10.3. Why $p$-adic space?}
On the classical level the main distinguishing feature of our model is the ultrametric $p$-adic structure 
of the classical mental space.  As we have already mentioned in section 3,  ulrametricity is simply 
a topological representation of hierarchy. Hence, the main classical feature of the model is its very 
special hierarchic structure. I think that the presence of such a hierarchic structure is the
very important condition of cognition and consciousness. In principle, it is possible to consider 
general ultrametric cognitive  models.  I restrict myself to  consideration of $p$-adic models,
since there is the possibility to connect $p$-adic hierarchic model with frequency domain models.

However,  the presence of the $p$-adic hierarchy
is not sufficient to induce consciousness (nor even cognition). For example, spin glasses
have hierarchic structures that in some cases could also be  mathematically described by 
$p$-adic numbers, see [62], [63]. The crucial point may be a complex system of interconnections between the
huge ensemble of hierarchic neural structures in brain.

{\bf 10.4. Individual and ensemble interpretations}
The large diversity of physical interpretations of the mathematical formalism of quantum 
mechanics is one of many serious problems in quantum foundations. Different interpretations
provide totally different views to physical reality (including the absence of such 
a reality at all), see e.g. [10]-[12], [15]. 

As a consequence of the great success of the books of R. Penrose
on the quantum approach to mind,  neurophysiologists, psychologists, cognitive scientists, and philosophers
are now quite familiar with one very special interpretation, namely Penrose's {\it quantum gravity}
improvement of the {\it conventional interpretation} of quantum mechanics. 

The first question is:
Why does the {\it conventional
interpretation} need some improvements at all?

This was well explained in book [15].\footnote{Neurophysiologists, psychologists, cognitive scientists, and philosophers
are lucky that R. Penrose does not support orthodox views to quantum theory. So in his books [14], [15]
this theory was not presented in the rigid orthodox form.}
Conventional quantum theory with the orthodox Copenhagen interpretation
has many problems including numerous mysteries and paradoxes 
(e.g. [15], p. 237: {\small ... yet it contains many mysteries. ... it provides us with a
very strange view of the world indeed."}; or R. Feynman: {\small ``It is all mysterious. And the 
more you look at it the more mysterious it seems."}). 

Unfortunately, all these mysteries and paradoxes related to the interpretation of quantum mechanics
were automatically transmited to cognitive sciences.
Some people enjoy this and they are happy to speak about mental nonlocality or mental collapse. 
It is the general attitude
to couple the mystery of consciousness with some still unclear aspects of interpretation 
of quantum mechanics.
On the other hand, many realistically thinking neurophysiologists, 
psychologists, cognitive scietists, and philosophers
dislike to use all such tricky quantum  things as superposition of (e.g. position) states for an individual system,
collapse, nonlocality, 
death of reality in the cognitive framework. I strongly suupport this viewpoint.\footnote{Of course,
discussing all these intriguing problems of interpretation of quantum mechanics we should not forget that
(as it was mentioned by one of referee's of this paper):
``The interpretation of the formalism is, in the end, largely irrelevant to the description of the phenomena
under observation or the mathematics used to treat them."}

There is no any possibility to go deeply into foundations of conventional quantum theory. I think that the crucial
point is the {\it individual interpretation} of a wave function. The wave function is associated with an individual 
quantum system (in the orthodox approach -- it gives the complete description). For example, the {\it individual interpretation}
induces such a mysterious thing as superposition (e.g. position) states for an individual system. 
On the other hand, individual superposition immediately implies that {\small ``Quantum theory provides a
superb description of physical reality on a small scale...",} [15], p.237. As the superposition of states for individual
macroscopic objects (e.g. cars) was never observed, conventional quantum theory should be applied on so called 
quantum scale. In particular, all cognitive models based on conventional quantum theory should go deeply beyond
the macroscopic neural level, see [15], p.355: {\small ``It is hard to see how one could usefully consider a quantum superposition
consisting of one neuron {\it firing,} and simultaneously {\it not firing.}"} Therefore all such models
suffer of the huge gap between quantum micro and neural macro scales. Of course, there are various 
attempts to solve these problem. For example, in [15] it was proposed to use quantum 
coherence to produce some macro states by coherence of large ensembles of quantum systems.

Finally, we mention the quantum gravity improvement of conventional quantum theory, [14], [15]. This is really an 
improvement and not a cardinal change of conventional quantum ideology. It is an attempt to explain
reduction as {\small ``gravitationally induced state-vector reduction."} It would not be useful to discuss
 the role of such an improvement  of physical theory in a biological journal. However, for cognitive models,
the use of quantum gravity arguments looks as just increasing of conventional quantum mystifcation. There
is a new huge gap
between quantum scale and Planck scale $(10^{-33}$ cm). It is even less belivable that the mind is 
induced by superpositions of mass states.

Quantum-like approach to cognitive modeling used in this paper is based on so called {\it ensemble
interpretation} of quantum mechanics, see e.g. L. Ballentine [11]. By this interpretation (that was strongly supported
by A. Einstein) a wave function is associated not with an individual physical system, but with a statistical
ensemble of systems. The statistical approach has its advantages and disadvantages. In particular, there is no
mystery of state reduction. Individual systems are not in superposition of different states. Superposition
of wave functions is a purely statistical property of various ensembles of physical systems. One of the main
problems of the statistical approach was the impossibility to get interference of probabilities on the 
basis of classical ensemble probability. Recently it was done in author's works [33]-[36] by taking into account
context dependence of probabilities. The absence of the mysterious
superposition for individual systems and operation with ensembles gives the possibility to apply the
Hilbert space probabilsitic formalism, quantum-like theory, to ensembles of macroscopic systems.
We agree with R. Penrose that an individual neuron could not be in superposition of two states,
but two ensembles of neurons (as well as
the same ensemble at distinct moments) could demonstrate features of superposition.

{\bf 10.5. Neural code and structure of mental space.}
Suppose that the coding system of a cognitive
system  is based on a frequency code. There exists an interval of  physical time $\Delta$ 
such that a classical mental state (mental position)  produced by a hierarchic chain of neurons
is a sequence with coordinates given by numbers of oscillations for corresponding neurons during
the interval $\Delta.$ This $\Delta$ depends on a cognitive system and even on a psychological function
inside the same brain, namely $\Delta= \Delta_{\tau, F}.$ Thus in our model the problem of the neural code is closely related to the problem
of time-scaling in neural systems. For different $\Delta,$ we get different coding systems, and, consequently,
different structures of mental spaces. The corresponding natural number $p$ that determines the $p$-adic 
structure on the mental space is defined as the maximal number of oscillations  that could be performed 
by neurons (in hierarchic chains of neurons working for some fixed psychological function) 
for the time interval $\Delta.$ The coding that is based on e.g. the 2-adic system induces the 2-adic mental space
that differs crucially from  the 5-adic (or 1997-adic) mental space induced by 
the 5-adic (or 1997-adic) system. As it was remarked in section 3, by changing the $p$-adic
structure we change crucially dynamics. Hence, the right choice of the time scaling parameter 
$\Delta$  and corresponding $p=p_{\Delta}$ plays the important role in the creation of an adequate mathematical model for
functioning of a psychological function.

{\bf 10.6. Mental time.}
There might be some connection between the time scale parameter
$\Delta$ of neural coding and {\it mental time.} There are strong experimental evidences,
see e.g. K. Mogi [66],  that a moment in mental time correlates with $\approx 100$ ms of physical time
for neural activity. In such a model the basic assumption is that the physical time required for
the transmission of information over synapses is somehow neglected in the mental time.
A moment in mental time is subserved by neural activities in different brain regions at different
physical times.
\medskip

{\bf 10.7. Quantum-like  models with $p$-adic valued functions.}
A series of works of the  author and his collaborators, see, for example, [47],
developed the formalism of quantum mechanics in which not only the classical
configuration space, but also wave functions are $p$-adic. Originally this formalism
was developed for high energy physics, namely, for theory of $p$-adic strings. Later I used this
formalism for cognitive modeling, see e.g. paper [29] on $p$-adic cognitive
pilot wave model (``conscious field model'') giving the very special realization of 
Bohm-Hiley-Pylkk\"anen ideas on active information. From the mathematical point of view
the $p$-adic valued formalism looks more attractive than the complex valued formalism
developed by Vladimirov and Volovich, see e.g. [45]. In particular, here operators of mental
position and motivation are well defined. However, there is a difficulty that induces 
strong prejudice against this $p$-adic valued formalism, namely the appearence of 
$p$-adic valued probablities. Despite very successful mathematical 
development of the theory with $p$-adic valued probabilities [47], it is clear that we cannot 
use it for ordinary measurements over physical and cognitive systems. In such measurements
we always observe ordinary probabilities. Thus $p$-adic valued quantum-like formalism 
could not be used to describe traditional mental measurements over a cognitive system
performed by external systems. As it was pointed out in [32], such $p$-adic
probabilities  (stabilization of frequencies in $p$-adic topology and chaotic 
behaviour of these frequencies in ordinary real topology)  might appear in anomalous phenomena. 
In principle, such probabilities might be related to the functioning of consciousness. It 
might be that consciousness uses self-measurements following to $p$-adic valued quantum-like
theory. However, in the present paper we would not like to study such a model of consciousness.

Finally, we mention the fundamental work of M. Pitk\"anen [67] that also contains a $p$-adic
model of consciousness. However, M. Pitk\"anen used an orthodox Copenhagen interpretation of quantum
mechanics (state reduction, superposition of states for individual quantum systems and so on).
Another fundamental aspect of his approach is the proposal to formulate space-time geometry
by using both real and $p$-adic space-time regions.

{\bf 10.8. Real and $p$-adic spaces.}
At first glance, in our model there is no direct connection between
real continuous space that is traditionally used to describe 
classical states of material objects and $p$-adic hierarchic (treelike)
spaces that was proposed to decribe classical mental states of brain.
So we follow to Descaurtes doctrine. Such an approach
was strongly critisized from many sides. In particular, such a theory 
is not coherent, see Whitehead [1]. Of course, it would be nice to develop
some classical and corresponding quantum-like models based on real/$p$-adic space.
The real and $p$-adic parts of material--mental space would describe the physical brain and 
``mental brain", respectively. Our first point  is that, in general, we could not work with the
fixed $p$-adic structure. As we have already discussed,  different cognitive systems
and psychological functions can be based on different $p$-adic mental spaces. Thus
in a general model we have to use all $p$-adic spaces simultaneously. We remark that a mathematical
topological structure unifying real and all $p$-adic numbers (for prime $p)$ is well known.
This is so called adelic space, see [45] on physical models over adels. The next natural step would be to apply 
adelic quantum-like formalism to measurements over material and cognitive systems.
In adelic quantum-like model ``the disastrous separation of body and mind, characteristic of philosophical systems which are in 
any important respect derived from Cartesianism" (see [1], p.348) could be avoided, since adelic amplitudes
would depend both on body (real) and mind ($p$-adic) variables.

{\bf 10.9. Microtubules.}
Are neurons really the basic elements for hierarchic mental coding? At the moment there 
is strong neurophysiological
evidences that this is really the case. Nevertheless, we should not totally reject other possibilities.
In particular, over the last 20 years S. Hameroff and his collaborators, see e.g. [68], have been developing
a model of consciousness based on quantum processes in microtubules.  Hameroff's approach
is a traditional quantum reductionst approach. Thus our paper has nothing to do with it.  However, the general idea
that microtubules play an important role in information processing in brain should be considered very seriously
in a quantum-like approach. Of course, in such a model the main role would be played by hierarchic
organization of microtubules on a classical level. Quantum-like formalism can be used to describe 
the corresponding mental amplitude. Finally, we mention some other fundamental papers on the
quantum brain [69]-[71].

{\bf 10.10. Non-reductionism.}
The basic question of all quantum reductionst models of consciousness is 
{\small ``How is it that consciousness can arise from such seemingly unpromising ingredients as 
matter, space, and time?} -- [15], p. 419. In our model, consciousness has no direct relation to matter.
It is a feature of very special hierarchic configuration of information described by the mental
amplitude $\phi(x).$ By answering to Penrose's question {\small ``The physical phenomenon of consciousness?",}
[15], p.406, I say: {\small ``Consciousness is  a bio-physical as well as a bio-information phenomenon."}

{\bf 10.11. Quantative measure of consciousness.} I was extremely facsinated by   Baars' idea to 
consider consciousness as a variable [72]. The main problem is to find some numerical representation
of such a consciousness-variable. In our model, such a variable should be in some way connected
with the basic probability distribution $P(t,x)= \vert \phi(t,x)\vert^2.$ 
This is the probability that the concrete hierarchic configuration of firings (e.g. configuration of frequencies) 
is realized in brain at the moment $t.$ Hence, if sufficiently many hierarchic chains of neurons
produce $x,$ then $P(t,x)$ is sufficiently large.  
The value of $P(t,x)$ by itself cannot be taken as a quantative measure of mentality.

For instance, suppose that $P(t,x)\equiv 1$ for all $x.$ This is the uniform distribution on 
the $p$-adic space. We could not expect that such an amplitude with uniform activation of all
classical mental states corresponds to a high level of mentality.\footnote{In particular, the free
mental wave $\phi_\xi(x)$ induces such a probability distribution. In such a state a cognitive system
has the fixed motivation $\xi.$ By proceeding with a fixed motivation (aim, task) a cognitive system $\tau$
performs not conscious, but merely AI-behaviour (for example, realization of a program given by the string 
of digits $\xi).$} Conscious behaviour corresponds to a mixture of various motivations. Such a mixture
is characterized by variation of the probability distribution $P(t,x).$ I propose the following
numerical measure of consciousness (at mental state $\phi(t,x)):$
$$
{\cal M}_{\rm{consciousness}}= \int_{{\bf Q}_p} (\vert D_x P(t,x)\vert^2 + 
\vert \frac{\partial P(t,x)}{\partial t} \vert^2 ) d x
$$

{\bf 10.12. Neural groups.} The fundamental role that internally organized groups of neurons (and not
individual neurons) play in processing of information in brain was discussed in details in Edelman's
theory of neural groups selection (TNGS), [73]. Our model in that  neural pathways are used 
as the neural (classical) basis for processing of information in brain is closely related to TNGS.
Of course, we understand that 
our model may be oversimplified. It may that the basic units should be  not chains, but
whole trees of neurons.

{\bf 10.13. Does consciousness benefit from long neural pathways?} Finally, we discuss one of the 
greatest mysteries of neuroanatomy, see, for example,  [17], [52]-[54], [73], [15]. It seems that in the process of neural 
evolution cognitive systems tried to create for each psychological function
neural pathways that are as long  as possible. This mystery might be explained on the basis of our neural
pathway coding model. If such a coding be really the case, then a cognitive system $\tau$ gets
great benefits by extending neural pathways for some psychological function as long as possible.
For example, let the neural code basis $p=5$ and a psychological function $F$ is based on very short
pathways of the length $L=2.$ Then the corresponding mental space contains $N(5,2)= 2^5= 32$ points.
Let now $p=5$ and $L=10000.$ Then the corresponding mental space contains huge number of 
points $N(5,10000)= 10^{20}$ points. On the latter (huge) mental space  mental amplitudes 
having essentially more complex behaviour (and, consequently, the measure of consciousness)  can be realized. 
It might be that
this mental space extending argument can be used to explain spatial separation of various maps in the brain,
see e.g. Edelman [73]. 

\medskip

\centerline{\bf Summary}

\medskip

{\small The crucial difference between my model and other classical as well as
quantum models of cognition is the use of {\it the $p$-adic
configuration space,} mental space, instead of a (traditionally used) {\it continuous real physical space.}
This is a kind of information model of brain (even on the `classical level'). However, we could not say
that there exist two totally different worlds -- mental and physical. Connection with physical 
space  is performed via the use of the special topology, namely the ulrametric $p$-adic
topology. In fact, the use of the concrete number system, $p$-adic numbers, is not so important.
It is just the simplest (but, of course, very important) model in that the special spatial structure,
namely hierarchic treelike structure (e.g., hierarchic neuronal structures that generate the hierarchic
frequency coding of cognitive information), is represented by an {\bf ultrametric
topological space.} Once we did such a transition from the physical space to an ultrametric space,
we can forget (at least at the first stage of modelling) about physical space and work 
in the ultrametric mental space. We understand well that at the moment
our fundamental postulate on ultrametric topology of information brain has only indirect
confirmations in neurophysiology and psychology, namely hierarchic processing of cognitive information,
see e.g. [17],[52]-[54],[72], [73]. In principle, on the basis of hierarchy we could reconstruct the
corresponding ultrametric topology. However, the problem of creating of the detailed map of hierarchic
structures in brain is far from its solution. Therefore, it is quite natural to try to start
simulation with the simplest ultrametric space -- the $p$-adic one.

The next crucial step is the use of the quantum-like formalism to describe mathematically the process of thinking in that
brain operates with {\bf probability distributions.} Here a mental state is mathematically described  by a probability
distribution.
This formalism is based on the standard quantum probabilistic 
calculus in a Hilbert space. However, in the opposite to orthodox quantum views, a quantum state (normalized
vector belonging to a Hilbert space) describes not an individual microscopic quantum system (e.g., an electron), 
but a statistical ensemble of macroscopic neuronal structures.\footnote{Thus our model differs crucially from 
quantum cognitive models making possible macroscopic and macrotermal quantum coherence, see e.g.  [67]. 
In our model, each individual neuron is (as everybody would agree) purely classical, but ensembles of neurons
could demonstrate quantum-like collective behaviour.}

We need not consider superposition of states for an individual 
neuronal system. Thus there is no problem with interpretation of such a superposition. Nevertheless,
a quantum-like brain is able to produce interference of mental states -- complex amplitudes of various
ensemble distributions. 

Our model can be called {\bf a model of probabilistic quantum-like thinking on the ultrametric space of 
hierarchic neuronal pathways.}}

\medskip

Main results of this paper were presented in author's talk and poster at the Conferences ``Toward a
Science of Consciousness" in Sh\"ovde, Sweden (2001), and Arizona (2002), [74].

\medskip

I would like to thank S. Albeverio, L. Ballentine, E. Beltrametti,
T. Hida, D. Greenberger, S. Gudder, I. Volovich, W. De Muynck, J. Summhammer, P. Lahti, J-A. Larsson, H. Atmanspacher, 
B. Coecke, S. Aerts, A. Peres, A. Holevo,  E. Loubenets,  L. Polley, A. Zeilinger, C. Fuchs, R. Gill, L. Hardy,
S. Goldshtein, A. Plotnitsky, A. Shimony, R. Jozsa, J. Bub, C. Caves, K. Gustafsson, H. Bernstein
for fruitful (and rather critical) discussions on the structure of quantum formalism.
I would like to thank D. Amit, B. Hiley, S. Greenfield, P. Pilkk\"anen, G. Vitiello
for discussions on cognitive modeling.

{\bf References}

1. A. N. Whitehead, 
{\it Process and Reality: An Essay in Cosmology.}  Macmillan Publishing Company, New York(1929).

2. A. N. Whitehead,  {\it  Adventures of Ideas.}  Cambridge Univ. Press, London (1933)

3. A. N. Whitehead, {\it Science in the modern world.} Penguin, London  (1939).

4. A. Shimony, 
On Mentality, Quantum Mechanics and the Actualization of Potentialities. 
In R. Penrose, M. Longair (Ed.) {\it The large, the small and the human mind.}
 Cambridge Univ. Press, New York(1997).

5. E. Schr\"odinger, {\it Philosophy and the Birth of Quantum Mechanics.}
Edited by M. Bitbol, O. Darrigol. Editions Frontieres (1992).

6. J. von Neumann, {\it Mathematical foundations
of quantum mechanics.} Princeton Univ. Press, Princeton, N.J. (1955).

7. W. Heisenberg, {\it  Physics and philosophy.}  Harper \& Row, Harper Torchbooks, New York(1958).

8. E.P. Wigner, The problem of measurement. {\it Am. J. Phys.}, {\bf 31}, 6 (1963);
{\it Symmetries and reflections.} Indiana Univ. Press, Bloomington (1967).

9. N. D. Mermin, Is the moon there when nobody looks? Reality and quantum theory.
{\it Phys. Today,} 38-41, April 1985.

10. A. Peres, {\it Quantum Theory: Concepts and Methods.} Kluwer Academic Publishers (1994).

11. L. E. Ballentine, {\it Quantum mechanics.} Englewood Cliffs, 
New Jersey (1989).

12. B. d'Espagnat, {\it Conceptual foundations of Quantum Mechanics.} Perseus Books, Reading,
Mass. (1999).

13. Y. F. Orlov,  The wave logic of consciousness: A hypothesis. 
{\it Int. J. Theor. Phys.} {\bf 21}, 1, 37-53 (1982) .

14. R. Penrose, {\it The emperor's new mind.} Oxford Univ. Press, New-York (1989).

15. R. Penrose  {\it Shadows of the mind.}  Oxford Univ. Press, Oxford (1994).

16 P. M. Churchland, {\it Matter and consciousness.} MIT Press, Cambridge (1999).

17. A.  Clark, {\it Psychological models and neural mechanisms. An
examination of reductionism in psychology.} Clarendon Press, Oxford (1980).

18. K. Lorenz, {\it On aggression.} Harcourt, Brace and World, New York
(1966).

19. B. F. Skinner, {\it Science and human  behaviour.} Macmillan Co.,
New York (1953).

20. R. Dawkins, { The selfish gene.} Oxford University Press, New York
(1976).

21. H. P. Stapp (1993) {\it Mind, matter and quantum mechanics.}  Springer-Verlag, Berlin-New York-Heidelberg.

22. M. Jibu, K. Yasue, A physical picture of Umezawa's quantum brain dynamics.
In {\it Cybernetics and Systems Research}, ed. R. Trappl (World Sc., London, 1992).

23. M. Jibu, K. Yasue, {\it Quantum brain dynamics and consciousness.}
J. Benjamins Publ. Company, Amsterdam/Philadelphia. 

24. H. Umezawa, {\it Advanced field theory: micro, macro, and thermal physics.}
American Inctitute of Physics, New-York (1993).

25. G. Vitiello, {\it My double unveiled - the dissipative quantum model of brain.}
J. Benjamins Publ. Company, Amsterdam/Philadelphia (2001).

26. D. Bohm,  and B. Hiley , {\it The undivided universe:
an ontological interpretation of quantum mechanics.}
Routledge and Kegan Paul, 
London (1993).

27. B. Hiley, P. Pylkk\"anen,  Active information and cognitive science -- 
A reply to Kiesepp\"a. In: Brain, mind and physics.  Editors: Pylkk\"anen, P., Pylkk\"o, P., Hautam\"aki, A..
IOS Press, Amsterdam (1997).

28. B. Hiley, Non-commutavive geometry, the Bohm interpretation and the mind-matter relationship.
To appear in Proc. CASYS 2000, LIege, Belgium, 2000.

29. A. Yu. Khrennikov,  Classical and quantum mechanics on $p$-adic trees of ideas.
{\it BioSystems,}  {\bf 56}, 95-120  (2000)

30. M. Lockwood, {\it Mind, Brain and Quantum.} Oxford,Blackwell (1989).

31. J. A. Barrett, {The quantum mechanics of minds and worlds.} Oxford Univ. Press, 1999.

32. A. Yu. Khrennikov,  Classical and quantum mechanics on information spaces
with applications to cognitive, psychological,
social and anomalous phenomena. {\it  Found. Phys.} {\bf 29,}  1065-1098 (1999).

33. A. Yu. Khrennikov, Origin of quantum probabilities. Proc. Conf.
"Foundations of Probability and Physics", {\it Quantum Probability and White Noise Analysis},
{\bf 13}, 180-200, WSP, Singapore (2001).

34. A. Yu. Khrennikov, Linear representations of probabilistic transformations induced
by context transitions. {\it J. Phys.A: Math. Gen.,} {\bf 34}, 9965-9981 (2001).

35. A. Khrennikov, Ensemble fluctuations and the origin of quantum probabilistic rule.
{\it J. Math. Phys.}, {\bf 43}, N. 2, 789-802 (2002).

36. A. Yu. Khrennikov, {\it Hyperbolic Quantum Mechanics.} Preprint: quant-ph/0101002, 31 Dec 2000.

37.  P. A. M.  Dirac, {\it The Principles of Quantum Mechanics}
Claredon Press, Oxford (1995).

38. W. Heisenberg, {\it Physical principles of quantum theory.}
(Chicago Univ. Press, 1930).

39. A. Yu. Khrennikov, Human subconscious as the p-adic dynamical 
system. {\it J. of Theor. Biology} {\bf 193}, 179-196(1998) .

40. A. Yu. Khrennikov,  $p$-adic dynamical systems: description of concurrent struggle in biological 
population with limited growth. {\it Dokl. Akad. Nauk} {\bf 361}, 752-754(1998).

41. S. Albeverio , A. Yu. Khrennikov, P. Kloeden, Human memory and
a $p$-adic dynamical systems. {\it Theor. and Math. Phys,} {\bf 117,} N.3, 385-396 (1998).

42. A. Yu. Khrennikov,  Description of the operation of the human subconscious by means 
of $p$-adic dynamical systems. {\it Dokl. Akad. Nauk} {\bf 365}, 458-460(1999).

43. D. Dubischar D., V. M. Gundlach, O. Steinkamp, 
A. Yu. Khrennikov, A $p$-adic model for the process of thinking 
disturbed by physiological and information noise. {\it J. Theor. Biology,}
{\bf 197,} 451-467 (1999).

44. A. Yu.  Khrennikov, 
$p$-adic discrete dynamical systems and collective behaviour of 
information states in cognitive models. {\it Discrete Dynamics in Nature and Society} {\bf 5,} 59-69 (2000).

45. V. S. Vladimirov, I. V. Volovich, E. I.    and Zelenov,
{\it $p$-adic Analysis and  Mathematical Physics}, World Scientific Publ.,
Singapore (1994).

46. I. V. Volovich,   $p$-adic string. {\it Class. Quant. Grav.},
{\bf 4}, 83--87  (1987).

47. A. Yu. Khrennikov,  Non-Archimedean analysis: quantum paradoxes, dynamical
systems and biological models. Kluwer Academic Publ.,
Dordrecht(1997).

48. L. Bianchi,  The functions of the frontal lobes. {\it{Brain,}} {\bf{18}}, 497-530(1895). 

49. I. P. Pavlov,  {\it Complete Works.} Academy of Science Press, Moscow  (1949).

50. W. Bechterew,  {\it Die Funktionen der Nervencentra.}
Fischer, Jena (1911).

51. H. Eichenbaum, R. A.  Clegg, and A. Feeley, 
Reexamination of functional subdivisions of the rodent prefrontal cortex. 
{\it Exper. Neurol.} {\bf{79}}, 434-451 (1983).

52. J. M. D. Fuster, 
The prefrontal cortex: anatomy, physiology, and neuropsychology of the frontal lobe. (1997)

53.  A. R. Damasio,  
{\it Descartes' error: emotion, reason, and the human brain.} Anton Books, New York(1994).

54. H. Damasio, A. R.  Damasio, {\it Lesion analysis in neuropsychology.}  Oxford Univ. Press, New-York  (1989).

55. I. Kant, {\it Critique of pure reason.} Macmillan Press, 1985.

56. D. J. Chalmers, 
{\it The conscious mind: in search of a fundamental theory.}  Oxford Univ. Press, New York(1996) .

57.  A. Yu. Khrennikov, {\it Supernalysis.} Nauka, Moscow,
(1997) (in
Russian). English translation: Kluwer Academic Publ.,  Dordreht (1999).

58. F. C. Hoppensteadt,  {\it  An introduction to the mathematics of neurons:
modeling in the frequency domain.}  Cambridge Univ. Press,
New York (1997).

59. F. C. Hoppensteadt  and E. Izhikevich,  Canonical models in
mathematical neuroscience. {\it  Proc. of Int. Math. Congress}, Berlin, 
{\bf 3}, 593-600 (1998).

60. A. J. Lemin, The  category of ultrametric spaces is isomorphic
to the category of complete, atomic, tree-like, and 
real graduated lattices LAT. {\it Algebra universalis,}, to be published.

61. M. Mezard, G. Parisi, M. Virasoro, {\it Spin-glass theory and beyond.}
World Sc., Singapure (1987).

62. G. Parisi, N. Sourlas, $p$-adic numbers and replica smmetry breaking.
{\it The European Physical J.}, {\bf 14B}, 535-542 (2000).

63. V. A. Avetisov, A. H. Bikulov, S. V. Kozyrev,
Application of $p$-adic analysis to models of breaking 
of replica symmetry. {\it J. Phys. A: Math. Gen.,} {\bf 32}, 8785-8791 (1999).

64.   L. Arnold,  {\it Random dynamical systems.} Springer Verlag, Berlin-New
York-Heidelberg (1998).

65. A. Khrennikov, {\it On cognitive experiments to test quantum-like behaviour 
of mind.} quant-ph/0205092 (2002).

66. K. Mori, On the relation between physical and psychological time.
Proc. Int. Conf. {\it Toward a
Science of Consciousness,} p. 102, Tucson, Arizona (2002).

67. M. Pitk\"anen, {\it TGD inspired theory of consciousness with applications to biosystems.}
Electronic vesrion: http://www.physics.helsinki.fi/~matpitka.

68. S. Hameroff, Quantum coherence in microtubules. 
A neural basis for emergent consciousness? {\it J. of Consciousness Studies,}
{\bf 1,} 91-118 (1994); Quantum computing in brain microtubules? The Penrose-Hameroff Orch Or model
of consciousness. {\it Phil. Tr. Royal Sc., London,} A, 1-28 (1998).

69. N. E.Mavromatos, D. V. Nanopoulos, I. Samaras, K. Zioutas,
{\it Advances in structural biology}, {\bf 5}, 127-134 (1998).

70. A. Mershin, D. V. Nanopoulos, E. M. C. Skoulakis, Quantum brain? Preprint
quant-ph/0007088 (2000).

71. A. Priel, N. Wolf, S. Hameroff, Dynamical properties of dendric arrays of microtubules - relevance to 
consciousness. Proc. Int. Conf. {\it Toward a
Science of Consciousness,} p. 69, Tucson, Arizona (2002).

72.  B. J. Baars,  {\it In the theater of consciousness. The workspace of mind.}
Oxford University Press, Oxford (1997).

73. G. M.  Edelman, {\it The remembered present: a biological theory of consciousness.}
New York, Basic Books, 1989.

74. A. Khrennikov, Quantum-like formalism for cognitive measurements.
Proc. Int. Conf. {\it Toward a
Science of Consciousness,} p. 272, Tucson, Arizona (2002).
\end{document}